\documentclass[hidelinks,a4,12pt]{article}

\usepackage{amssymb}
\usepackage{amsmath}
\usepackage{amscd}
\usepackage{mathrsfs}
\usepackage{amssymb}
\usepackage{feynmf}
\usepackage[pdftex]{graphicx}
\usepackage[pdftex,bookmarks=true,pdfauthor={},pdftitle={},pagebackref=false]{hyperref}
\usepackage{youngtab}
\usepackage{cite}
\usepackage{color}
\usepackage{mdframed}
\usepackage{breakurl}
\usepackage{slashed,bbold}
\usepackage{hyperref} 
\hypersetup{colorlinks,urlcolor=blue,citecolor=blue,linkcolor=blue}

\numberwithin{equation}{section}

\allowdisplaybreaks

\begin{document}

\setlength{\oddsidemargin}{0cm}
\setlength{\baselineskip}{7mm}

\thispagestyle{empty}
\setcounter{page}{0}

\begin{flushright}

\end{flushright}

\vspace*{-1cm}

\begin{center}
{\bf \Large }
\end{center}

\begin{center}
{\bf \Large High Temperature Massless Scalar Partition Function  
\\[0.4cm]
on General Stationary Backgrounds}


\vspace*{1cm}

Manuel Valle$^{a,}$\footnote{\tt  manuel.valle@ehu.es}
and Miguel \'A. V\'azquez-Mozo$^{b,}$\footnote{\tt 
vazquez@usal.es}

\end{center}

\vspace*{0.0cm}

\begin{center}

$^{a}${\sl EHU Quantum Center \\ 
Universidad del Pa\'{\i}s Vasco/Euskal Herriko Unibertsitatea EHU \\
48080 Leioa, Spain
}

\vspace*{0.2cm}

$^{b}${\sl Departamento de F\'{\i}sica Fundamental \\
 Universidad de Salamanca \\ 
 Plaza de la Merced s/n, 37008 Salamanca, Spain
  }
\end{center}

\vspace*{0.7cm}

\centerline{\bf \large Abstract}

\noindent
The high-temperature equilibrium partition function of a massless real scalar field 
nonminimally coupled to the scalar curvature is computed at second order in the derivative expansion on a generic stationary
background. Using covariant perturbation theory, 
the
expression of the thermal partition function at
second order in powers of curvatures is also obtained, including its nonlocal contributions. 
For conformal coupling, the Weyl anomaly
at fourth order in derivatives and second order in curvatures is evaluated 
using both expansions and the results found to be consistent.

\newpage  

\setcounter{footnote}{0}

\tableofcontents

\section{Introduction}

Since its inception in the 
1970s~\cite{Deser:1976yx,Duff:1977ay,Duff:1993wm,Deser:1996na}, 
the Weyl anomaly
has played a central role in physics with
many realizations in various fields~\cite{Mukhanov:2007zz,Nakayama:2013is}.
As with other types of anomalies~\cite{Bertlmann:1996xk,Alvarez-Gaume:2022aak}, 
quantum violations of scale invariance also have 
implications for hydrodynamics and transport theory, some of which have been 
investigated in a number of works (see, for 
example,~\cite{Eling:2013bj,Chernodub:2016lbo,Diles:2017azi,Romatschke:2017ejr,Chernodub:2020dqo}).
Many of these analyses rely on computing the theory's one-loop effective action 
on a generic stationary background, in terms of which 
the equilibrium partition function at one loop is obtained~\cite{Jensen:2012jh,Banerjee:2012iz},
as well as the constitutive relations for the different currents and the associated
transport coefficients.

The local part of the high-temperature partition function for a scalar field on 
static or stationary geometries 
has been the subject of attention in the literature 
(see, for example,~\cite{Dowker:1978md,Dowker:1988jw,Gusev:1997ns,Gusev:1998rp,Kalinichenko:2013fj}).
Nonlocal contributions, on the other hand, have been evaluated so far only on ultrastatic 
backgrounds~\cite{Gusev:1997ns,Gusev:1998rp}. The task set in the present work is precisely to obtain 
an {\em explicit} expression of the finite-temperature equilibrium partition function for a 
massless scalar field on a {\em generic} stationary 
background and arbitrary nonminimal coupling to background curvature, 
including also the nonlocal terms.

Apart from its immediate intrinsic technical interest, 
the generalization of the massless scalar high-temperature partition function 
to arbitrary stationary metrics has interesting applications to the 
study of the hydrodynamics of scalar fluids, and the r\^ole of the Weyl anomaly in this context.
Comparing with the ultrastatic case 
studied so far in the literature, the expressions obtained here using covariant perturbation theory
include the contributions of the gradients of four additional 
metric functions, related to fluid vorticity, acceleration, and
spatial changes in the local Tolman temperature.

The plan of this paper is as follows.
In Section~\ref{sec:setup} we present the calculation of the scalar
high-temperature partition function on a general stationary geometry 
to second order in the derivative expansion, as well as
of the associated Weyl anomaly. In order to 
analyze nonlocal contributions, in Section~\ref{sec:nonlocal} we carry out an expansion to
second order in curvatures of the partition function, finding the existence of extra
form factors associated with fluid vorticity and local temperature gradients. 
We finish in Section~\ref{sec:closing} with a brief
discussion of our results and prospects for some future work.  

\section{Equilibrium scalar partition function to second order in derivatives}
\label{sec:setup}

We consider a 
massless scalar field~$\varphi(x)\equiv\varphi(t,\mathbf{x})$
on a curved spacetime with action
\begin{align}
S={1\over 2}\int d^{4}x\sqrt{-G}\,\varphi\big(\Box-\xi\mathscr{R}\big)\varphi,
\label{eq:action_scalar}
\end{align}
where~$\Box$ is the four-dimensional Laplace-Beltrami 
operator,~$\mathscr{R}$ the four-dimensional
scalar curvature, and~$\xi\in\mathbb{R}$ the nonminimal coupling constant. 
We moreover
assume a generic stationary background described by the line element 
\begin{align}
ds^{2}&=G_{\mu\nu}(\mathbf{x})dx^{\mu}dx^{\nu} \nonumber \\[0.2cm]
&=-e^{2\sigma(\mathbf{x})}\big[dt+a_{i}(\mathbf{x})dx^{i}\big]^{2}
+g_{ij}(\mathbf{x})dx^{i}dx^{j}.
\label{eq:stationary_line_element}
\end{align}
In the context of hydrodynamics, 
the metric functions~$\sigma(\mathbf{x})$ and~$a_{i}(\mathbf{x})$
are related to the fluid velocity, vorticity, acceleration, 
and local (Tolman) temperature,
while the transverse metric~$g_{ij}(\mathbf{x})$ is the source of the 
energy-momentum current~\cite{Banerjee:2012iz}. The ultrastatic case considered 
in refs.~\cite{Gusev:1997ns,Gusev:1998rp} corresponds to setting~$\sigma(\mathbf{x})=a_{i}(\mathbf{x})=0$.

Using the  
imaginary time formalism~\cite{Kapusta:2006pm},
the finite-temperature equilibrium partition function is computed as
\begin{align}
\log Z(\beta_{0})&={1\over 2}\sum_{n\in\mathbb{Z}}{\rm tr\,}\log\big(-\Box+\xi\mathscr{R}\big)
\nonumber \\[0.2cm]
&=-\left.{1\over 2}{\partial\over\partial\epsilon}\left[{\mu^{2\epsilon}\over\Gamma(\epsilon)}
\sum_{n\in\mathbb{Z}}\int_{0}^{\infty}{ds\over s^{1-\epsilon}}{\rm tr\,}
e^{s(\Box-\xi\mathscr{R})}\right]\right|_{\epsilon=0},
\label{eq:Z(beta)_generic}
\end{align}
where the sum is over bosonic Matsubara 
frequencies~$\omega_{n}={2\pi n\over \beta_{0}}$ (with~$n\in\mathbb{Z}$ 
and~$\beta_{0}$ the period of the Euclidean time coordinate) and 
a regularization scale~$\mu$ has been introduced.
The four-dimensional Laplace-Beltrami
operator~$\Box$ acting on Fourier-transformed fields~$\varphi(\omega_{n},\mathbf{x})$ reads
\begin{align}
\Box &={1\over \sqrt{g}}\partial_{i}\big(\sqrt{g}g^{ij}\partial_{j}\big)
-\big(2\omega_{n}a^{i}-\partial_{i}\sigma g^{ij}\big)\partial_{j}
-\omega_{n}a^{j}\partial_{i}\sigma \nonumber \\[0.2cm]
&+\omega_{n}^{2}\big(a^{i}a_{i}-e^{-2\sigma}\big)
-{\omega_{n}\over\sqrt{g}}\partial_{i}\big(\sqrt{g}a^{i}\big),
\label{eq:box_general}
\end{align}
whereas the four-dimensional scalar curvature~$\mathscr{R}$ of the
stationary metric~\eqref{eq:stationary_line_element} can be written as
\begin{align}
\mathscr{R}&=R+{1\over 4}e^{2\sigma}f_{ij}f^{ij}-2g^{ij}\partial_{i}\sigma\partial_{j}\sigma
-{2\over \sqrt{g}}\partial_{i}\big(\sqrt{g}g^{ij}\partial_{j}\sigma\big),
\end{align}
with~$R$ the three-dimensional 
scalar curvature associated with the transverse metric~$g_{ij}$,
which is the one used to raise and lower spatial indices, and~$f_{ij}=\partial_{i}a_{j}
-\partial_{j}a_{i}$ the field strength of the Kaluza-Klein (KK) gauge field.
In fact, the operator~$L\equiv -\big(\Box-\xi\mathscr{R}\big)$ appearing in 
Eq.~\eqref{eq:Z(beta)_generic} 
can be recast in the more suggestive form
\begin{align}
L&=-{1\over \sqrt{g}}\big(\partial_{i}+\mathcal{A}_{i}\big)
\Big[\sqrt{g}g^{ij}\big(\partial_{j}+\mathcal{A}_{j}\big)\Big]+\mathcal{Q} \nonumber \\[0.2cm]
&\equiv -\triangle_{\mathcal{A}}+\mathcal{Q},
\label{eq:suggestive_form_Box}
\end{align}
where the generalized 
connection~$\mathcal{A}_{i}$ and potential~$\mathcal{Q}$ are given by
\begin{align}
\mathcal{A}_{i}&=-\omega_{n}a_{i}+{1\over 2}\partial_{i}\sigma, \nonumber \\[0.2cm]
\mathcal{Q}&=e^{-2\sigma}\omega_{n}^{2}
+\xi R+{\xi\over 4}e^{2\sigma}f_{ij}f^{ij}
\label{eq:def_A_Q} \\[0.2cm]
&-\left(2\xi-{1\over 4}\right)g^{ij}\partial_{i}\sigma
\partial_{j}\sigma
-\left(2\xi-{1\over 2}\right)\triangle\sigma.
\nonumber 
\end{align}
Here, we have also introduced a generalized covariant 
derivative~$\nabla_{i}^{\mathcal{A}}\equiv \nabla_{i}+\mathcal{A}_{i}$,
such that~$\triangle_{\mathcal{A}}\equiv g^{ij}\nabla_{i}^{\mathcal{A}}\nabla_{j}^{\mathcal{A}}$
is its corresponding Laplace-Beltrami operator,
and denoted~$\triangle\equiv g^{ij}\nabla_{i}\nabla_{j}$. 
Notice as well that the field strength 
of the generalized connection
\begin{align}
\mathcal{F}_{ij}&=\partial_{i}\mathcal{A}_{j}
-\partial_{j}\mathcal{A}_{i} \nonumber \\[0.2cm]
&=-\omega_{n}f_{ij},
\end{align}
determines the commutator of
the covariant derivatives acting on
scalar fields
\begin{align}
[\nabla_{i}^{\mathcal{A}},\nabla_{j}^{\mathcal{A}}]\varphi&=\mathcal{F}_{ij}\varphi.
\end{align}
This relation will be important when applying the results of ref.~\cite{Barvinsky:1990up}
to the calculation presented in the next section.

The partition function~\eqref{eq:Z(beta)_generic} can then be written
as
\begin{align}
\log Z(\beta)=-{1\over 2}{\partial\over\partial\epsilon}\left[\left.
{\mu^{2\epsilon}\over\Gamma(\epsilon)}\sum_{n\in\mathbb{Z}}\int_{0}^{\infty}{ds\over s^{1-\epsilon}}
\int d^{3}\mathbf{x}\sqrt{g}\,
\mathcal{K}(\omega_{n};\mathbf{x},\mathbf{x};s)\right]\right|_{\epsilon=0},
\label{eq:logZ(beta)_heat_kernel}
\end{align}
where~$\mathcal{K}(\omega_{n};\mathbf{x},\mathbf{x}';s)$ is the heat kernel of the 
operator~\eqref{eq:suggestive_form_Box},
solving the equation
\begin{align}
\left({\partial\over\partial s}+L\right)\mathcal{K}(\omega_{n};\mathbf{x},\mathbf{x}';s)=0
\hspace*{0.7cm} \mbox{with}\hspace*{0.7cm}
\lim_{s\rightarrow 0}\mathcal{K}(\omega_{n};\mathbf{x},\mathbf{x}';s)
={1\over \sqrt{g(\mathbf{x})}}\delta^{(3)}(\mathbf{x}-\mathbf{x}').
\end{align} 
The derivative expansion of the partition function
is implemented by the Seeley-DeWitt series of the heat kernel in the coincidence limit
\begin{align}
\mathcal{K}(\omega_{n};\mathbf{x},\mathbf{x};s)
\sim (4\pi s)^{-{3\over 2}}e^{-s(\mathcal{Q}-{1\over 6}R)}
\sum_{k=0}^{\infty}{(-1)^{k}s^{k}\over k!}b_{k}(\mathbf{x}),
\label{eq:shortpropertimeexp_general}
\end{align}
where the first few coefficients are given 
by~\cite{Gilkey:1975iq,Avramidi:1990je,Avramidi:2000bm,Vassilevich:2003xt,Avra2015}
\begin{align}
b_{0}(\mathbf{x})&=1, \nonumber \\[0.2cm]
b_{1}(\mathbf{x})&=0, \nonumber \\[0.2cm]
b_{2}(\mathbf{x})&=-{1\over 3}\triangle\mathcal{Q}
+{1\over 6}\mathcal{F}_{ij}\mathcal{F}^{ij}
+{1\over 15}\triangle R
-{1\over 90}R_{ij}R^{ij}+{1\over 90}R_{ijk\ell}R^{ijk\ell}, \nonumber \\[0.2cm]
b_{3}(\mathbf{x})&=-{1\over 2}g^{ij}\partial_{i}\mathcal{Q}\partial_{j}\mathcal{Q}
+{1\over 10}\triangle^{2}\mathcal{Q}
-{2\over 15}g^{k\ell}\nabla_{k}f_{ij}\nabla_{\ell}f^{ij}
-{1\over 30}\nabla_{j}f_{i}^{\,\,\,j}\nabla_{k}f^{ik}
\label{eq:bs} \\[0.2cm]
&-{1\over 5}f_{ij}\triangle f^{ij}
+{1\over 15}R^{ij}\nabla_{i}\nabla_{j}\mathcal{Q}
+{1\over 5}g^{ij}\partial_{i}\mathcal{Q}\partial_{j}R
+{1\over 5}f_{ij}f^{jk}f_{k}^{\,\,\,i} \nonumber \\[0.2cm]
&+{1\over 10}f^{ij}R_{ijk\ell}f^{k\ell}
-{1\over 15}f^{ij}R_{jk}f^{k}_{\,\,\,i}
+\ldots
\nonumber
\end{align}
Here,~$R_{ijk\ell}$ and~$R_{ij}$ are, respectively, the Riemann and Ricci tensors
of the transverse metric~$g_{ij}$, while
in the expression of~$b_{3}(\mathbf{x})$ the ellipsis stands for terms with more than four derivatives. 
As shown in 
ref.~\cite{Parker:1984dj} (see also~\cite{Parker:1985mv,Parker:2009uva}), 
the exponential in Eq.~\eqref{eq:shortpropertimeexp_general}
resums nonderivative contributions of~$\mathcal{Q}$ and~$R$ to all orders in~$s$,
which are therefore absent from the coefficients shown in~\eqref{eq:bs}.

A calculation of the partition
function~\eqref{eq:logZ(beta)_heat_kernel} using eqs.~\eqref{eq:shortpropertimeexp_general}
and~\eqref{eq:bs} shows that, to second order in derivatives, only the 
coefficients~$b_{2}(\mathbf{x})$ and~$b_{3}(\mathbf{x})$ contribute, leading to the 
result\footnote{In writing this equation, we exploited the 
freedom associated with the existence of the total derivative 
combination~$\beta^{-1}\big[\partial_{i}(\sqrt{g}g^{ij}\partial_{j}\sigma)-\sqrt{g}g^{ij}\partial_{i}\sigma
\partial_{j}\sigma\big]=\partial_{i}\big(\beta^{-1}\sqrt{g}g^{ij}\partial_{j}\sigma\big)$,
taking the 
terms~$\beta^{-1}R$,~$\beta^{-1}\sqrt{g}g^{ij}\partial_{i}\sigma
\partial_{j}\sigma$, and~$f_{ij}f^{ij}$ as independent.}
\begin{align}
\log Z(\beta)&=\int d^{3}\mathbf{x}\sqrt{g}\left[
-{\pi^{2}\over 90\beta^{3}}+{1-6\xi\over 144\beta}\Big(4g^{ij}\partial_{i}\sigma\partial_{j}\sigma
-R\Big)+{1+6\xi\over 576\beta}e^{2\sigma}f_{ij}f^{ij}
+\ldots\right],
\label{eq:partition_function_gen}
\end{align}
with~$\beta(\mathbf{x})=\beta_{0}e^{\sigma(\mathbf{x})}$ the inverse local equilibrium temperature
and the ellipsis indicates higher-order terms in the derivative expansion.
Notice that the partition function 
is explicitly invariant under KK gauge transformations,~$a_{i}\rightarrow a_{i}(\mathbf{x})
+\partial_{i}\chi(\mathbf{x})$, corresponding to 
position-dependent time shifts~$t\rightarrow t-\chi(\mathbf{x})$ in the background 
metric~\eqref{eq:stationary_line_element}.
For
conformal coupling,~$\xi={1\over 6}$, the second term in~\eqref{eq:partition_function_gen}
vanishes and the thermal partition function to this
quadratic derivative order
remains invariant under Weyl 
rescalings~$G_{\mu\nu}(\mathbf{x})\rightarrow e^{2\lambda(\mathbf{x})}G_{\mu\nu}(\mathbf{x})$, 
acting as
\begin{align}
\sigma(\mathbf{x}) &\longrightarrow \sigma(\mathbf{x})+\lambda(\mathbf{x}), \nonumber \\[0.2cm]
\beta(\mathbf{x}) &\longrightarrow e^{\lambda(\mathbf{x})}\beta(\mathbf{x}), 
\label{eq:Weyltrans}\\[0.2cm] 
g_{ij}(\mathbf{x})&\longrightarrow e^{2\lambda(\mathbf{x})}g_{ij}(\mathbf{x}), \nonumber
\end{align}
while the KK field~$a_{i}(\mathbf{x})$ and its field strength~$f_{ij}(\mathbf{x})$ do not transform. 
Incidentally, the results for ultrastatic backgrounds in ref.~\cite{Gusev:1998rp}
are retrieved by setting~$\sigma(\mathbf{x})=a_{i}(\mathbf{x})=0$. 
Equation~\eqref{eq:partition_function_gen} also generalizes the
partition function obtained in ref.~\cite{Fursaev:2000dv} for~$\xi=0$ to arbitrary~$\xi$.

For any theory, whether conformal or not, the coefficients of the 
terms~$\log(\beta\mu)$ in the partition function originate from
the contributions in the heat kernel expansion~\eqref{eq:shortpropertimeexp_general} 
of order~$s^{1/2}$,~$s^{3/2}\omega_{n}^{2}$, 
and~$s^{5/2}\omega_{n}^{4}$. Using the Seeley-DeWitt coefficients
given in Eq.~\eqref{eq:bs}, we find the following logarithmic dependence
on the renormalization scale~$\mu$ 
for a generic stationary background\footnote{In fact, together with the expressions given in Eq.~\eqref{eq:bs}, 
we also need 
those four-derivative terms in~$b_{4}(\mathbf{x})$ multiplying~$\omega_{n}^{4}$, 
which read~${3\over 5}(\nabla e^{-2\sigma})^{2}+{4\over 5}g^{ij}\partial_{i}e^{-2\sigma}
\partial_{j}\triangle e^{-2\sigma}$ (see refs.~\cite{Avramidi:1990ug,Avramidi:2000bm,Avra2015}).}
\begin{align}
\mu{\partial\over\partial\mu}\log Z(\beta)&=
-{1\over 32\pi^{2}}\int d^{3}\mathbf{x}\sqrt{g}\beta\left[
{1\over 90}\triangle R-{1\over 90}R_{ij}R^{ij}+{1\over 90}R_{ijk\ell}R^{ijk\ell}
+{1\over 3}\triangle \overline{P}
\right.\nonumber\\[0.2cm]
&+{1\over 45}e^{2\sigma}\nabla_{k}f_{ij}\nabla^{k}f^{ij}
+{1\over 180}e^{2\sigma}\nabla_{j}f^{ij}\nabla_{k}f_{i}^{\,\,\,k}
+{1\over 30}e^{2\sigma}f_{ij}\triangle f^{ij} \nonumber \\[0.2cm]
&-{1\over 60}e^{2\sigma}\triangle^{2}e^{-2\sigma}-{1\over 6}e^{2\sigma}g^{ij}\partial_{i}e^{-2\sigma}
\partial_{j}\overline{P}-{1\over 180}e^{2\sigma}g^{ij}\partial_{i}e^{-2\sigma}\partial_{j}R
\label{eq:Weyl_anomaly_full}\\[0.2cm]
&-{1\over 90}e^{2\sigma}R^{ij}\nabla_{i}\nabla_{j}e^{-2\sigma}
+{3\over 80}e^{4\sigma}(\triangle e^{-2\sigma})^{2}
+{1\over 20}e^{4\sigma}g^{ij}\partial_{i}e^{-2\sigma}\partial_{j}\triangle e^{-2\sigma}
\nonumber \\[0.2cm]
&-\left.{1\over 6}e^{2\sigma}\overline{P}\triangle e^{-2\sigma}
+{1\over 12}e^{2\sigma}\overline{P}f_{ij}f^{ij}
+{1\over 8}e^{4\sigma}\overline{P}g^{ij}\partial_{i}e^{-2\sigma}\partial_{j}e^{-2\sigma}
+\overline{P}^{2}
\right],
\nonumber 
\end{align}
where, to simplify, 
we introduced the notation
\begin{align}
\overline{P}
&\equiv-\left(\xi-{1\over 6}\right)R-{\xi\over 4}e^{2\sigma}f_{ij}f^{ij}+\left(2\xi-{1\over 4}\right)
g^{ij}\partial_{i}\sigma\partial_{j}\sigma 
+\left(2\xi-{1\over 2}\right)\triangle\sigma
\label{eq:Pbar_def} \\[0.2cm]
&=-\left(\xi-{1\over 6}\right)R
-{\xi\over 4}e^{2\sigma}f_{ij}f^{ij} 
-{1\over 2}\left(3\xi-{5\over 8}\right)
e^{4\sigma} g^{ij}\partial_{i}e^{-2\sigma}\partial_{j}e^{-2\sigma}
+{1\over 2}\left(2\xi-{1\over 2}\right)e^{2\sigma}\triangle e^{-2\sigma}, 
\nonumber
\end{align}
that will also be used in the next section. 
All terms in~\eqref{eq:Weyl_anomaly_full} are of fourth order in the derivative expansion.
It can be checked that each term 
inside the square bracket on the right-hand side of~\eqref{eq:Weyl_anomaly_full} 
scales as~$e^{-4\lambda}$
under {\em rigid} Weyl transformations~\eqref{eq:Weyltrans},
thus canceling the change
of the common factor~$\sqrt{g}\beta\rightarrow e^{4\lambda}\sqrt{g}\beta$ and leaving
the whole expression invariant.

For conformal coupling,~$\xi=1/6$, Eq.~\eqref{eq:Weyl_anomaly_full} 
renders the expression of the scale anomaly and must include the square
of the Weyl tensor, as it behooves 
a type-B Weyl anomaly~\cite{Deser:1993yx,Deser:1996na}.
Its presence, in fact, can be read off Eq.~\eqref{eq:Weyl_anomaly_full}. 
Indeed, in perturbation theory
the nonlocal (Euclidean) zero-temperature effective action on a four-dimensional 
stationary backgrounds takes the form~(cf. refs.~\cite{Deser:1976yx,Deser:1993yx,Deser:1996na,
Karateev:2023mrb})
\begin{align}
W_{\rm anom}=c\int d^{3}x\sqrt{g(\mathbf{x})}\,\beta(\mathbf{x})\, \mathscr{C}_{\mu\nu\alpha\beta}(\mathbf{x})
\log\left({\mu^{2}\over -\triangle}\right)\mathscr{C}^{\mu\nu\alpha\beta}(\mathbf{x}) + \ldots,
\label{eq:W_anomaly}
\end{align}
where~$\mathscr{C}^{\mu}_{\,\,\,\,\nu\alpha\beta}$ is the four-dimensional 
Weyl tensor,~$c$ the trace 
anomaly, and~$\sqrt{g}\beta\equiv \int_{0}^{\beta_{0}}d\tau
e^{\sigma}\sqrt{g}$ 
comes from the integration over the compact 
Euclidean time. In this expression, 
we have stressed that all quantities in the integrand and independent of the 
time coordinate and
the ellipsis stands for  terms of cubic order in the 
curvatures~$R$,~$R_{ij}$,~$f_{ij}$, and $\overline{P}$, as well as possibly some 
terms proportional to~$\Box R$ or $R^{2}$.
Under a Weyl rescaling~\eqref{eq:Weyltrans}, the anomalous effective action~\eqref{eq:W_anomaly}
transforms as
\begin{align}
\delta_{\lambda}W_{\rm anom}=2 c\int d^{3}x\,\sqrt{g(\mathbf{x})}
\lambda(\mathbf{x})\beta(\mathbf{x})\, 
\mathscr{C}_{\mu\nu\alpha\beta}(\mathbf{x})\mathscr{C}^{\mu\nu\alpha\beta}(\mathbf{x}),
\label{eq:Weyl_anomaly_gen}
\end{align} 
rendering the expression of the type-B Weyl anomaly.
 
At finite temperature, on the other hand, 
after summing over Matsubara frequencies 
the logarithmic part of 
the effective action 
exhibits the following structure\footnote{As it will be discussed in the next section, the logarithmic Laplacian
appearing in the zero-temperature term~\eqref{eq:W_anomaly} is canceled by a similar contribution from
the finite-temperature piece.}
\begin{align}
W_{\rm anom}=\int d^{3}x\sqrt{g(\mathbf{x})}\,\beta(\mathbf{x})\sum_{i,j}\mathfrak{R}_{i}(\mathbf{x})
\chi_{ij}\mathfrak{R}_{j}(\mathbf{x})\log
\left[{\beta(\mathbf{x})\mu\over 4\pi}\right],
\end{align} 
where~$\mathfrak{R}_{i}$ represents combinations of metric functions of second order in derivatives
and~$\chi_{ij}$ are appropriate differential form factors. 
This is heuristically seen by considering the
simpler example of a stationary metric with~$\sigma=0$ and~$g_{ij}=\delta_{ij}$. Then, we have
\begin{align}
{1\over 2}\int d^{4}x\sqrt{-G}\,\mathscr{C}_{\mu\nu\alpha\beta}\mathscr{C}^{\mu\nu\alpha\beta}
&=\int d^{4}x\sqrt{-G}\left(\mathscr{R}_{\mu\nu}\mathscr{R}^{\mu\nu}-{1\over 3}\mathscr{R}^{2}\right)
\nonumber \\[0.2cm]
&={\beta_{0}\over 2}\int d^{3}x \partial_{j}f^{ij}\partial_{k}f_{i}^{\,\,k} \\[0.2cm]
&=-{\beta_{0}\over 4}\int d^{3}x f_{ij}\triangle f^{ij},
\nonumber
\end{align}
where in the last line we have carried out various integrations by parts and dropped
total derivative terms, with~$\beta_{0}$ arising from the integration over the
Euclidean compact time~$\tau$. When properly normalized, this reproduces the
combined three terms in the second
line of Eq.~\eqref{eq:Weyl_anomaly_full} with~$\sigma=0$ and~$g_{ij}=\delta_{ij}$, 
up to boundary contributions.

\section{Nonlocal terms}
\label{sec:nonlocal}

Our analysis so far has been based on expanding the high-temperature equilibrium
partition function in the number of derivatives. In order to 
study nonlocal terms, on the other hand, we should rather consider an expansion in 
powers of the curvatures following the analysis of~\cite{Barvinsky:1990up} 
(see also~\cite{Barvinsky:1990uq,Barvinsky:1994ic,Barvinsky:1994cg,Barvinsky:1992ts}).
In order to apply the techniques presented in these references, we rewrite the
operator~\eqref{eq:suggestive_form_Box} in the form
\begin{align}
L&=-\left(\triangle_{\mathcal{A}}-{1\over 6}R\right)-\omega_{n}^{2}+P,
\label{eq:L_op_def}
\end{align}
where we introduced
\begin{align}
\label{eq:PP}
P&\equiv \omega_{n}^{2}\big(1-e^{-2\sigma}\big)+\overline{P},
\end{align}
and~$\overline{P}$, defined in Eq.~\eqref{eq:Pbar_def}, 
is independent of the Matsubara frequencies. 
The constant term~$-\omega_n^2$ in~\eqref{eq:L_op_def}
gives rise to a prefactor~$e^{-\omega_n^2 s}$ in the heat kernel, 
while the position-dependent term~$\omega_{n}^{2}e^{-2\sigma}$ is included in the definition of~$P$. 
The methods and expressions of~\cite{Barvinsky:1990up} can now be applied to 
write the following solution to the heat kernel equation of~$L$ in powers of the
curvatures~$R$,~$R_{ij},~\mathcal{F}_{ij}$, and the potential~$P$
\begin{align}
\int d^{3}\mathbf{x}\,\sqrt{g}\sum_{n\in\mathbf{Z}}\mathcal{K}(\omega_{n};\mathbf{x},\mathbf{x};s)
&=(4\pi s)^{-{3\over 2}}\int d^{3}\mathbf{x}\,\sqrt{g}\sum_{n\in\mathbb{Z}}e^{-s\omega_{n}^{2}}
\Bigg\{1+sP \nonumber \\[0.2cm]
&+s^{2}\Bigg[R_{ij}f_{1}(-s\triangle_{\mathcal{A}})R^{ij}
+Rf_{2}(-s\triangle_{\mathcal{A}})R
+Pf_{3}(-s\triangle_{\mathcal{A}})R \label{eq:nonlocal_heatkernel_first}\\[0.2cm]
&+Pf_{4}(-s\triangle_{\mathcal{A}})P
+\mathcal{F}_{ij}f_{5}(-s\triangle_{\mathcal{A}})\mathcal{F}^{ij}
\Bigg]+\ldots\Bigg\},
\nonumber
\end{align}
where the ellipsis indicates terms of third and higher order, that we drop\footnote{As shown 
in~\cite{Barvinsky:1990up}, the Riemann tensor~$R_{ijk\ell}$ can be eliminated 
inside the integral in terms of~$R$ and~$R_{ij}$.}.
The differential operators~$f_{i}(-s\triangle_{\mathcal{A}})$ are defined by the integrals
\begin{align}
f_{1}(-s\triangle_{\mathcal{A}})&={1\over 12}\int_{0}^{1}d\alpha\,\alpha^{4} 
\exp\left({1-\alpha^{2}\over 4}s\triangle_{\mathcal{A}}\right), \nonumber
\\[0.2cm]
f_{4}(-s\triangle_{\mathcal{A}})&={1\over 2}\int_{0}^{1}d\alpha\,
\exp\left({1-\alpha^{2}\over 4}s\triangle_{\mathcal{A}}\right),  \\[0.2cm]
f_{5}(-s\triangle_{\mathcal{A}})&={1\over 4}\int_{0}^{1}d\alpha\,\alpha^{2}
\exp\left({1-\alpha^{2}\over 4}s\triangle_{\mathcal{A}}\right),\nonumber
\end{align}
together with the relations
\begin{align}
f_{2}(-s\triangle_{\mathcal{A}})&={1\over 144}f_{4}(-s\triangle_{\mathcal{A}})
-{1\over 12}f_{5}(-s\triangle_{\mathcal{A}})-{1\over 8}f_{1}(-s\triangle_{\mathcal{A}}), \nonumber \\[0.2cm]
f_{3}(-s\triangle_{\mathcal{A}})&={1\over 6}f_{4}(-s\triangle_{\mathcal{A}})
-f_{5}(-s\triangle_{\mathcal{A}}).
\end{align}
Since  the various form factors  
$f_i(-s \triangle_{\mathcal{A}})$
are connecting terms of first order in curvatures, 
in extracting the contribution of second order 
one can replace~$\triangle_{\mathcal{A}}$ with 
$\triangle$.

The equilibrium partition function is now obtained plugging 
the expansion~\eqref{eq:nonlocal_heatkernel_first} 
into~\eqref{eq:logZ(beta)_heat_kernel},
which reads
\begin{align}
\log Z(\beta)
&=\int d^{3}\mathbf{x}\,\sqrt{g}\left[-{\pi^{2}\over 90\beta_{0}^{3}}
+{\pi^{2}\over 30\beta_{0}^{3}}\sigma-{\pi^{2}\over 30\beta_{0}^{3}}\sigma^{2}
+{6\xi-1\over 144\beta_{0}}R-{8\xi-1\over 96\beta_{0}}g^{ij}\partial_{i}\sigma\partial_{j}\sigma
\right.
\nonumber \\[0.2cm]
&+{\xi\over 96\beta_{0}}f_{ij}f^{ij}
+R_{ij}\chi_{1}\left(\beta_{0}\sqrt{-\triangle}\right)R^{ij}
+R\chi_{2}\left(\beta_{0}\sqrt{-\triangle}\right)R
\nonumber \\[0.2cm]
&+\overline{P}\chi_{3}\left(\beta_{0}\sqrt{-\triangle}\right)R
+\overline{P}\chi_{4}\left(\beta_{0}\sqrt{-\triangle}\right)\overline{P}
+f_{ij}\chi_{5}\left(\beta_{0}\sqrt{-\triangle}\right)f^{ij}
\label{eq:locallogZ}
\\[0.2cm]
&+2\sigma\chi_{6}(\beta_{0}\sqrt{-\triangle})R
+4\sigma\chi_{7}(\beta_{0}\sqrt{-\triangle})\overline{P}
+4\sigma\chi_{8}(\beta_{0}\sqrt{-\triangle})\sigma\bigg],
\nonumber
\end{align}
where we kept only terms that are bilinears in curvatures 
including~$P$, 
and expanded~$\beta=\beta_{0}e^{\sigma}$ to 
second order in~$\sigma$.  
This expansion is mandatory because~$\sigma$ has to be
consider first order in curvatures, 
as follows from~$P-\overline{P}=\omega_{n}^{2}(1-e^{-2 \sigma})\approx 2\omega_{n}^{2}\sigma$ 
[see Eq.~\eqref{eq:PP}]. 

The corresponding form factors are given by
\begin{align}
\chi_{i}(\beta_{0}\sqrt{-\triangle})&=
-{1\over 2}{\partial\over\partial\epsilon}\left.\left[{\mu^{2\epsilon}\over\Gamma(\epsilon)}
\int_{0}^{\infty}ds\,s^{1+\epsilon}(4\pi s)^{-{3\over 2}}\int d^{3}\mathbf{x}\,
\sqrt{g}\sum_{n\in\mathbb{Z}}e^{-s\omega_{n}^{2}}f_{i}(-s\triangle)
\right]\right|_{\epsilon=0}, 
\end{align}
for~$i=1,\ldots,4$ and
\begin{align}
\chi_{5}(\beta_{0}\sqrt{-\triangle})&=
-{1\over 2}{\partial\over\partial\epsilon}\left.\left[{\mu^{2\epsilon}\over\Gamma(\epsilon)}
\int_{0}^{\infty}ds\,s^{1+\epsilon}(4\pi s)^{-{3\over 2}}\int d^{3}\mathbf{x}\,
\sqrt{g}\sum_{n\in\mathbb{Z}}\omega_{n}^{2}e^{-s\omega_{n}^{2}}f_{5}(-s\triangle)
\right]\right|_{\epsilon=0},
\nonumber \\[0.2cm]
\chi_{6}(\beta_{0}\sqrt{-\triangle})&=
-{1\over 2}{\partial\over\partial\epsilon}\left.\left[{\mu^{2\epsilon}\over\Gamma(\epsilon)}
\int_{0}^{\infty}ds\,s^{1+\epsilon}(4\pi s)^{-{3\over 2}}\int d^{3}\mathbf{x}\,
\sqrt{g}\sum_{n\in\mathbb{Z}}\omega_{n}^{2}e^{-s\omega_{n}^{2}}f_{3}(-s\triangle)
\right]\right|_{\epsilon=0}, \label{eq:chi_i_general} \\[0.2cm]
\chi_{7}(\beta_{0}\sqrt{-\triangle})&=
-{1\over 2}{\partial\over\partial\epsilon}\left.\left[{\mu^{2\epsilon}\over\Gamma(\epsilon)}
\int_{0}^{\infty}ds\,s^{1+\epsilon}(4\pi s)^{-{3\over 2}}\int d^{3}\mathbf{x}\,
\sqrt{g}\sum_{n\in\mathbb{Z}}\omega_{n}^{2}e^{-s\omega_{n}^{2}}f_{4}(-s\triangle)
\right]\right|_{\epsilon=0}, \nonumber \\[0.2cm]
\chi_{8}(\beta_{0}\sqrt{-\triangle})&=
-{1\over 2}{\partial\over\partial\epsilon}\left.\left[{\mu^{2\epsilon}\over\Gamma(\epsilon)}
\int_{0}^{\infty}ds\,s^{1+\epsilon}(4\pi s)^{-{3\over 2}}\int d^{3}\mathbf{x}\,
\sqrt{g}\sum_{n\in\mathbb{Z}}\omega_{n}^{4}e^{-s\omega_{n}^{2}}f_{4}(-s\triangle)
\right]\right|_{\epsilon=0}.
\nonumber
\end{align}
Moreover, in writing the first line of~\eqref{eq:locallogZ} we dropped 
total derivative terms in~$P$ that do not contribute
to the partition function.
Notice that the form factors~$\chi_{5}$-$\chi_{7}$ 
originate from those terms in~$P$ proportional to~$\omega_n^2$, whereas~$\chi_8$ 
comes from the ones scaling as~$\omega_n^4$. 

To carry out the sums and integrations in the definitions of the form
factors, we apply Poisson resummation to the Matsubara series
\begin{align}
\sum_{n\in\mathbb{Z}}e^{-s\omega_{n}^{2}}
&=
{\beta_{0}\over \sqrt{4\pi s}}
\left(1+2\sum_{n=1}^{\infty}e^{-{n^{2}\beta^{2}\over 4s}}\right), \nonumber \\[0.2cm]
\sum_{n\in\mathbb{Z}}\omega_{n}^{2}e^{-s\omega_{n}^{2}}&=
{\beta_{0}\over 4\sqrt{\pi}s^{3\over 2}}
+{\beta_{0}\over 2\sqrt{\pi}s^{3\over 2}}\sum_{n=1}^{\infty}\left(1
-{n^{2}\beta_{0}^{2}\over 2s}\right)e^{-{n^{2}\beta_{0}^{2}\over 4s}}, \\[0.2cm]
\sum_{n\in\mathbb{Z}}\omega_{n}^{4}e^{-s\omega_{n}^{2}}
&={3\beta_{0}\over 8\sqrt{\pi}s^{5\over 2}}
+{3\beta_{0}\over 4\sqrt{\pi} s^{5\over 2}}\left(
\sum_{n=1}^{\infty}e^{-{n^{2}\beta_{0}^{2}\over 4s}}
-{\beta_{0}^{2}\over s}\sum_{n=1}^{\infty}n^{2}e^{-{n^{2}\beta_{0}^{2}\over 4s}}
+{\beta_{0}^{4}\over 4s^{2}}\sum_{n=1}^{\infty}n^{4}e^{-{n^{2}\beta_{0}^{2}\over 4s}}
\right),
\nonumber
\end{align}
after which the integration over~$s$ can be done in terms of modified Bessel functions which, in turn,
are expressed using exponential integrals (see ref.~\cite{Gusev:1998rp} for a somewhat 
similar calculation). After
some lengthy manipulations, we find the following results for all eight thermal form factors
depending on $z \equiv \beta_0 \sqrt{-\triangle}$
\begin{align}
\chi_{1}(z)&=-{\beta_{0}\over 512 z}
-{\beta_{0}\over 960\pi^{2}}\left[\gamma_{E}+\log\left({\beta_{0}\mu\over 4\pi}\right)\right]
+{\beta_{0}\zeta(3)z^{2}\over 107520\pi^{4}}
-{\beta_{0}\zeta(5)z^{4}\over 5160960\pi^{6}}
+{\beta_{0}\zeta(7)z^{6}\over 181665792\pi^{8}}+\mathcal{O}(z^{8}), \nonumber \\[0.2cm]
\chi_{2}(z)&={25\beta_{0}\over 36 864 z}
+{\beta_{0}\over 2880\pi^{2}}\left[\gamma_{E}+\log\left({\beta_{0}\mu\over 4\pi}\right)\right]
-{\beta_{0}\zeta(3)z^{2}\over 483840\pi^{4}} 
+{\beta_{0}\zeta(7)z^{6}\over 544997376\pi^{8}}
+\mathcal{O}(z^{8}), \nonumber \\[0.2cm]
\chi_{3}(z)&={\beta_{0}\over 384 z}+{\beta_{0}\zeta(3)z^{2}\over 23040\pi^{4}}
-{\beta_{0}\zeta(5)z^{4}\over 430080\pi^{4}}+{\beta_{0}\zeta(7)z^{6}\over 8257536\pi^{8}}+\mathcal{O}(z^{8}), \label{eq:form_factors}\\[0.2cm]
\chi_{4}(z)&=-{\beta_{0}\over 32 z}-{\beta_{0}\over 32\pi^{2}}\left[\gamma_{E}
+\log\left({\beta_{0}\mu\over 4\pi}\right)\right]+{\beta_{0}\zeta(3)z^{2}\over 1536\pi^{2}}
-{\beta_{0}\zeta(5)z^{4}\over 40960\pi^{6}}+{\beta_{0}\zeta(7)z^{6}\over 917504\pi^{8}}+\mathcal{O}(z^{8}), 
\nonumber \\[0.2cm]
\chi_{5}(z)&={1\over 576\beta_{0}}
+{z^{2}\over 3840\pi^{2}\beta_{0}}\left[1+\gamma_{E}+\log\left({\beta_{0}\mu\over 4\pi}\right)\right]
-{\zeta(3)z^{4}\over 143360\pi^{4}\beta_{0}}
+{\zeta(5)z^{6}\over 4128768\pi^{6}\beta_{0}}+\mathcal{O}(z^{8}),
\nonumber \\[0.2cm]
\chi_{6}(z)&={z^{2}\over 5760\pi^{2}\beta_{0}}\left[
1+\gamma_{E}+\log\left({\beta_{0}\mu\over 4\pi}\right)\right]
-{\zeta(3)z^{4}\over 107520\pi^{4}\beta_{0}}+{\zeta(5)z^{6}\over 2064384\pi^{6}\beta_{0}}
+\mathcal{O}(z^{8}), \nonumber \\[0.2cm]
\chi_{7}(z)
&={1\over 96\beta_{0}}+{z^{2}\over 384\pi^{2}\beta_{0}}\left[
1+\gamma_{E}+\log\left({\beta_{0}\mu\over 4\pi}\right)\right]
-{\zeta(3)z^{4}\over 10240\pi^{4}\beta_{0}}+{\zeta(5)z^{6}\over 229376\pi^{6}\beta_{0}}+\mathcal{O}(z^{8}), 
\nonumber \\[0.2cm]
\chi_{8}(z)&=-{\pi^{2}\over 240\beta_{0}^{3}}-{z^{2}\over 1152\beta_{0}^{3}}
-{z^{4}\over 7680\pi^{2}\beta_{0}^{3}}\left[4+3\gamma_{E}+3\log\left({\beta_{0}\mu\over 4\pi}\right)\right]
+{\zeta(3)z^{6}\over 57344\pi^{4}\beta_{0}^{3}}+\mathcal{O}(z^{8}),
\nonumber
\end{align}
with~$\gamma_{E}$ the Euler-Mascheroni constant. Higher corrections
can be computed systematically using the technique described to arbitrary order
in~$z$.
It should be pointed out that 
all contributions proportional to~$\log{z}$ cancel out among the vacuum and 
finite-temperature contributions, so the only nonlocal terms remaining are simple poles in~$z$.

It is important to stress at this point the fundamental difference between the 
expansion~\eqref{eq:locallogZ} in
powers of the curvatures~$R$,~$R_{ij}$,~$f_{ij}$, and~$\overline{P}$
and the one in the number of derivatives
leading to Eq.~\eqref{eq:partition_function_gen}. In fact, the two expressions cannot be compared
straightforwardly.
As an example, the coefficient of the term proportional to~$f_{ij}f^{ij}$ in 
Eq.~\eqref{eq:partition_function_gen} results from adding to the corresponding contribution
in the second line of Eq.~\eqref{eq:locallogZ}
the leading local piece of~$f_{ij}\chi_{5}(\beta_{0}\sqrt{-\triangle})f^{ij}$ 
in~\eqref{eq:form_factors}
\begin{align}
{\xi\over 96\beta_{0}}f_{ij}f^{ij}+{1\over 576\beta_{0}}f_{ij}f^{ij}
={1+6\xi\over 576\beta_{0}}f_{ij}f^{ij}.
\end{align}
Similarly, the part proportional to~$\sigma^{2}$ 
from the leading term in~$4\sigma\chi_{8}(\beta_{0}\sqrt{-\triangle})\sigma$
combines with the powers
of~$\sigma$ in the first line of~\eqref{eq:locallogZ}, to give
\begin{align}
-{\pi^{2}\over 90\beta_{0}^{3}}+{\pi^{2}\over 30\beta_{0}^{3}}\sigma
-{\pi^{2}\over 30\beta_{0}^{3}}\sigma^{2}-{\pi^{2}\over 60\beta_{0}^{3}}\sigma^{2}
&=-{\pi^{2}\over 90\beta_{0}^{3}}+{\pi^{2}\over 30\beta_{0}^{3}}\sigma
-{\pi^{2}\over 20\beta_{0}^{3}}\sigma^{2}.
\end{align}
This coincides with the expansion of 
the first term in the partition function~\eqref{eq:partition_function_gen} 
in powers of~$\sigma$ to quadratic order
\begin{align}
-{\pi^{2}\over 90\beta_{0}^{3}}e^{-3\sigma}
=-{\pi^{2}\over 90\beta_{0}^{3}}+{\pi^{2}\over 30\beta_{0}^{3}}\sigma
-{\pi^{2}\over 20\beta_{0}^{3}}\sigma^{2}+\ldots
\end{align}
As for gradients of~$\sigma$ and the curvature term, collecting the 
relevant contributions in~\eqref{eq:locallogZ}, we find, modulo total derivatives
\begin{align}
{6\xi-1\over 144\beta_{0}}R&-{8\xi-1\over 96\beta_{0}}g^{ij}\partial_{i}\sigma\partial_{j}\sigma
+4\sigma\chi_{7}(\beta_{0}\sqrt{-\triangle})\overline{P}
+4\sigma\chi_{8}(\beta_{0}\sqrt{-\triangle})\sigma\nonumber \\[0.2cm]
&={6\xi-1\over 144\beta_{0}}(1-\sigma)R-{6\xi-1\over 36\beta_{0}}g^{ij}\partial_{i}\sigma\partial_{j}\sigma
+\ldots,
\end{align}
reproducing the expansion in powers of~$\sigma$ of the 
local inverse temperature in Eq.~\eqref{eq:partition_function_gen}
\begin{align}
{1-6\xi\over 144\beta}\Big(4g^{ij}\partial_{i}\sigma\partial_{j}\sigma-R\Big)
={6\xi-1\over 144\beta_{0}}(1-\sigma)R-{6\xi-1\over 36\beta_{0}}g^{ij}\partial_{i}\sigma\partial_{j}\sigma
+\ldots
\end{align}
This match of the expansions~\eqref{eq:partition_function_gen} 
and~\eqref{eq:locallogZ} provides a nontrivial test of our results.
Accounting for higher orders in~$\sigma$ requires including contributions beyond those
in Eq.~\eqref{eq:nonlocal_heatkernel_first}.
 
A second important test is provided by the computation
of
the coefficients of the logarithmic terms in the form factors~\eqref{eq:form_factors},
giving the result
\begin{align}
\mu{\partial\over\partial\mu}\log Z(\beta)&=
-{\beta_{0}\over 32\pi^{2}}\int d^{3}\mathbf{x}\sqrt{g}\left(
{1\over 30}R_{ij}R^{ij}
+{1-20\xi+60\xi^{2}\over 60}R^{2}
+{1\over 120}f_{ij}\triangle f^{ij} 
\right.\nonumber\\[0.2cm]
&\left.
+{7-60\xi+120\xi^{2}\over 30}\sigma\triangle R
+{6-60\xi+180\xi^{2}\over 45}\sigma\triangle^{2}\sigma
\right),
\label{eq:Weyl_anomaly2}
\end{align}
where we have substituted the expression of~$\overline{P}$ given in Eq.~\eqref{eq:Pbar_def},
keeping only those terms bilinear in curvatures [see the remarks after Eq.~\eqref{eq:locallogZ}].
Their apparent differences notwithstanding, this expression is in full agreement with the general 
result 
provided in 
Eq.~\eqref{eq:Weyl_anomaly_full}. 
Indeed, expanding there all exponentials to linear 
order in~$\sigma$ and applying 
some Riemann and Ricci tensor identities, we retrieve the expression~\eqref{eq:Weyl_anomaly2}  
after
a number of integration by parts, retaining only those terms up to second order in curvatures.
For conformal coupling $\xi=1/6$, we obtain
the Weyl anomaly
\begin{align}
\mathcal{A}_{\rm Weyl}&=-{\beta_{0}\over 1920 \pi^{2}} \int d^{3}x \sqrt{g}\left[  
2\Big(R_{ij}R^{ij}-{1\over 3} R^2 \Big)
+{1\over 2}f_{ij}\triangle f^{ij} 
+{2\over 3} \sigma \triangle R 
+{4\over 3}\sigma\triangle^{2}\sigma  
\right],
\label{eq:Weyl_anomaly_final}
\end{align}
where the quantity between the square brackets 
is the four-dimensional Weyl tensor squared of
the stationary metric~\eqref{eq:stationary_line_element} and we omitted total derivative terms. 

A comparison of~\eqref{eq:locallogZ} 
with the result for ultrastatic backgrounds obtained in~\cite{Gusev:1998rp} shows the 
existence of four additional form factors~$\chi_{5}$-$\chi_{8}$ acting on~$\sigma$,~$R$ 
 and~$f_{ij}$. 
Interestingly, none of them give rise to
nonlocal terms, which are still associated solely with~$\chi_{1}$-$\chi_{4}$
\begin{align}
\log Z(\beta)_{\rm nonlocal}
&=-{1\over 32}\int d^{3}\mathbf{x}\sqrt{g}\left[
{1\over 16}R_{ij}{1\over \sqrt{-\triangle}}R^{ij}
-{25\over 1152}R{1\over \sqrt{-\triangle}}R \right.\nonumber \\[0.2cm]
&-\left.{1\over 12}\overline{P}{1\over \sqrt{-\triangle}}R
+\overline{P}{1\over \sqrt{-\triangle}}\overline{P}
\right].
\label{eq:nonlocal_terms}
\end{align}
Despite replicating the structure found in~\cite{Gusev:1998rp}, 
these nonlocal terms
incorporate through~$\overline{P}$ 
the effect of Tolman temperature gradients,~$T^{-1}\partial_{i}T=-\partial_{i}\sigma$,
as well as of the fluid vorticity encoded in the~KK 
field strength,~$\omega^{i}\sim {1\over 2}\epsilon^{ijk}f_{jk}$, none of them present 
for 
ultrastatic metrics.
The new form factors, however, do contribute to the Weyl anomaly, inducing the 
last three terms 
on the right-hand side of Eq.~\eqref{eq:Weyl_anomaly_final}.
 
Finally, let us point out that 
the structure of the nonlocal terms~\eqref{eq:nonlocal_terms} 
we obtained using functional methods can also be 
recovered from a calculation of three scalar one-loop diagrams at finite temperature
where we, respectively, insert
two gravitons~$h_{\mu\nu}$, two external sources~$J$ coupling to~$\varphi^{2}$, 
and one graviton~$h_{\mu\nu}$ and one
external source~$J$ (cf. refs.~\cite{Yamada:1974ea,Teixeira:2020kew}).
All these three diagrams contain contributions scaling as~$|\mathbf{p}|^{-1}$, with~$\mathbf{p}$ the
incoming three-momentum, corresponding to the four nonlocal terms shown in Eq.~\eqref{eq:nonlocal_terms}.
No nonlocal terms arise, however, from tadpole diagrams.

\section{Closing remarks}
\label{sec:closing}

In this paper we have computed the high-temperature 
equilibrium partition function of a massless scalar field on a general
stationary geometry nonminimally
coupled to the background scalar curvature, using expansions 
in the number of derivatives as well as in powers of curvatures. 
The results found in both cases display a dependence on 
local temperature gradients and fluid vorticity, encoded in the metric 
functions~$\sigma(\mathbf{x})$ and~$a_{i}(\mathbf{x})$, thus generalizing the results 
found for ultrastatic metrics to general stationary geometries and arbitrary curvature coupling. 
The general expression of the Weyl anomaly at
second order in curvatures has also been obtained from the 
logarithmic terms in both forms of the partition function. 
In a hydrodynamical context, our expressions can be used
to construct the fluid equilibrium partition function at finite temperature, including the effects
of the local temperature gradients, vorticity, and acceleration.

These implications of our results for hydrodynamics and transport are
one of the various interesting problems waiting to be addressed. 
Another concerns the extension of these results to physical 
situations in which conformal invariance 
is spontaneously broken~\cite{Schwimmer:2010za,Schwimmer:2023nzk}, as well as the possible 
application of the techniques introduced in ref.~\cite{Manes:2018llx} 
to incorporate the effect of 
spontaneous gauge
symmetry breaking to the case of the Weyl anomaly.
These and other questions will be investigated elsewhere.

\section*{Acknowledgments} 

The work of 
M. V. is supported by Spanish Science Ministry Grants PID2021-123703NB-C21 
(MCIU/AEI/FEDER, EU) and PID2024-156016NB-I00 (MCIU/AEI/FEDER, EU).
M. A. V.-M. acknowledges financial support from Spanish Science Ministry Grants
PID2021-123703NB-C22 (MCIU/AEI/FEDER, EU) and PID2024-160856NB-I00 (MCIU/AEI/FEDER, EU).
Both authors are also funded by Basque Government Grant IT979-16.


\end{document}